\UseRawInputEncoding

\documentclass[preprintnumbers,amsmath,amssymb,prb,
twocolumn,superscriptaddress]{revtex4}

\usepackage{graphicx}   
\usepackage{dcolumn}    
\usepackage{bm}         

\begin{document}

\title{Microwave spectroscopy of spin-orbit coupled states: energy detuning versus interdot coupling modulation}

\author{G. Giavaras}
\affiliation{Faculty of Pure and Applied Sciences, University of
Tsukuba, Tsukuba 305-8571, Japan}

\author{Yasuhiro Tokura}
\affiliation{Faculty of Pure and Applied Sciences, University of
Tsukuba, Tsukuba 305-8571, Japan} \affiliation{Tsukuba Research
Center for Energy Materials Science (TREMS), Tsukuba 305-8571,
Japan}


\begin{abstract}
We study the AC field induced current peaks of a spin blockaded
double quantum dot with spin-orbit interaction. The AC field
modulates either the interdot tunnel coupling or the energy
detuning, and we choose the AC field frequency range to induce two
singlet-triplet transitions giving rise to two current peaks. We
show that for a large detuning the two current peaks can be
significantly stronger when the AC field modulates the tunnel
coupling, thus making the detection of the spin-orbit gap more
efficient. We also demonstrate the importance of the time
dependence of the spin-orbit interaction.
\end{abstract}

\maketitle

\section{Introduction}

The singlet-triplet states of two electron spins in tunnel-coupled
quantum dots can be used to define spin-qubits in semiconductor
devices.~\cite{petta, morton} In the presence of a strong
spin-orbit interaction (SOI) an applied AC electric field can give
rise to singlet-triplet transitions and spin resonance can be
achieved.~\cite{ono2017, perge} In a double quantum dot which is
tuned to the spin blockade regime,~\cite{ono02} the transitions
can be probed by the AC induced current peaks. It has been
experimentally demonstrated that the two-spin energy spectra can
be extracted by examining the magnetic field dependent position of
the current peaks.~\cite{ono2017, perge} The exchange energy, the
strength of the SOI, as well as the $g$-factors of the quantum
dots can then be estimated. Microwave spectroscopy has also been
performed for the investigation of charge qubits,~\cite{hayashi,
fitch} as well as other hybrid spin systems.~\cite{wang, yang,
song} Charge localization in quantum dot systems can be controlled
with AC fields,~\cite{paspalakis, terzis, cref} while various
important parameters of the spin and/or charge dynamics can be
extracted from AC induced interference patterns.~\cite{shevche10,
shevche12, gallego, chatte}

In this work, we study the current through a double dot (DD) for
two different cases of the AC electric field; in the first case
the AC field modulates the interdot tunnel coupling of the DD, and
in the second case the AC field modulates the energy detuning of
the DD. We consider a specific energy configuration and AC
frequency range which involve two SOI coupled singlet-triplet
states, and a third state with (mostly) triplet character. The two
SOI-coupled singlet-triplet states form an anticrossing
point,\cite{ono2017, perge, takahashi, kanai} and in this work we
focus on this point.

Specifically, we perform electronic transport calculations and
demonstrate that for a large energy detuning the tunnel coupling
modulation results in stronger AC-induced current peaks than the
corresponding peaks induced by the detuning modulation. The
stronger peaks can offer a significant advantage when the
spectroscopy of the coupled spin system is performed by monitoring
the magnetic field dependence of the position of the peaks. When
the peaks are suppressed no reliable information can be extracted.

The tunnel coupling modulation offers a similar advantage when the
transitions involve only the two states forming the anticrossing
point.~\cite{giavaras19a} This finding together with the results
of the present work demonstrate that modulating the tunnel
coupling of a DD with an AC electric field is a robust method to
perform spectroscopy of spin-orbit coupled states. Furthermore, in
the present work, we explore the time dependent role of the SOI,
and specify the regime in which this time dependence should be
taken into account because it can drastically affect the
AC-induced current peaks.

In some experimental works the interdot tunnel coupling has been
accurately controlled and transport measurements have been
performed.~\cite{oosterkamp, bertrand} For instance, Bertrand
\textit{et al} [Ref.~\onlinecite{bertrand}] have demonstrated that
the tunnel coupling can be tuned by orders of magnitude on the
nanosecond time scale. Therefore, our theoretical findings could
be tested with existing semiconductor technology.

\section{Double quantum dot model}

We focus on the spin blockade regime~\cite{ono02} for two serially
tunnel-coupled quantum dots. In this regime the quantum dot 1 (dot
2) is coupled to the left (right) metallic lead and with the
appropriate bias voltage current can flow through the DD when the
blockade is partially or completely lifted. Each quantum dot has a
single orbital level and dot 2 is lower in energy by an amount
equal to the charging energy which is assumed to be much larger
than the tunnel coupling. Consequently, for the appropriate bias
voltage a single electron can be localized in dot 2 during the
electronic transport process.~\cite{ono02} If we use the notation
$(n, m)$ to indicate $n$ electrons on the dot 1 and $m$ electrons
on the dot 2 then electronic transport process through the DD
takes place via the charge cycle: $(0, 1)\rightarrow (1,
1)\rightarrow (0, 2)\rightarrow (0, 1)$. For the DD system there
are in total six two-electron states but in the spin blockade
regime the double occupation on dot 1 can be ignored because it
lies much higher in energy and does not affect the dynamics.
Therefore, the relevant two-electron states are the triplet states
$|T_+\rangle =
c^{\dagger}_{1\uparrow}c^{\dagger}_{2\uparrow}|0\rangle$,
$|T_-\rangle = c^{\dagger}_{1\downarrow}
c^{\dagger}_{2\downarrow}|0\rangle$, $|T_{0}\rangle=(
c^{\dagger}_{1\uparrow}c^{\dagger}_{2\downarrow}
+c^{\dagger}_{1\downarrow}c^{\dagger}_{2\uparrow})|0\rangle/\sqrt{2}$
and the two singlet states $|S_{02}\rangle =
c^{\dagger}_{2\uparrow}c^{\dagger}_{2\downarrow}|0\rangle$,
$|S_{11}\rangle=( c^{\dagger}_{1\uparrow}c^{\dagger}_{2\downarrow}
-c^{\dagger}_{1\downarrow}c^{\dagger}_{2\uparrow})|0\rangle/\sqrt{2}$.
The fermionic operator $c^{\dagger}_{i\sigma}$ creates an electron
on dot $i=1$, 2 with spin $\sigma=\uparrow$, $\downarrow$, and
$|0\rangle$ denotes the vacuum state. In this singlet-triplet
basis the DD Hamiltonian~\cite{giavaras19a} is
\begin{equation}
\begin{split}
&H_{\mathrm{DD}}=\Delta[|T_-\rangle\langle T_-|-|T_+\rangle\langle
T_+|]-\delta|S_{02}\rangle\langle
S_{02}|\\
&-\sqrt{2}T_{\mathrm{c}}|S_{11}\rangle\langle
S_{02}|+\Delta^{-}|S_{11}\rangle\langle T_{0}| + \text{H.c.}\\
&-T_{\mathrm{so}}[|T_+\rangle\langle S_{02}|+ |T_-\rangle\langle
S_{02}|] + \text{H.c.}\\
\end{split}
\end{equation}
Here, $\delta$ is the energy detuning, $T_{\mathrm{c}}$ is the
tunnel coupling between the two dots, and $T_{\mathrm{so}}$ is the
SOI-induced tunnel coupling causing a spin-flip.~\cite{mireles,
pan} The magnetic field is denoted by $B$ which gives rise to the
Zeeman splitting $\Delta_{i}=g_i \mu_{\mathrm{B}} B$ ($i=1$, 2) in
each quantum dot. Then $\Delta=(\Delta_{2}+\Delta_{1})/2$, and the
Zeeman asymmetry is $\Delta^{-}=(\Delta_{2}-\Delta_{1})/2$. To a
good approximation, in the transport process of a spin-blockaded
DD only the $c^{\dagger}_{2\uparrow}|0\rangle$,
$c^{\dagger}_{2\downarrow}|0\rangle$ single electron states are
important, and $H_{\mathrm{DD}}$ can be also derived using a
standard two-site Hubbard model.~\cite{giavaras13}

In the present work, we consider two cases for the AC field.
Specifically, in the first case the AC field modulates the energy
detuning of the DD, thus we consider the following time
dependence:
\begin{equation}\label{detuning}
\delta(t) = \varepsilon + A_{\mathrm{d}} \sin( 2\pi f t),
\end{equation}
where $A_{\mathrm{d}}$ is the AC amplitude and $f$ is the AC
frequency. The constant value of the detuning is denoted by
$\varepsilon$. In the second case, the AC field modulates the
interdot tunnel coupling, thus we assume the time dependent terms
\begin{equation}\label{barrier}
\begin{split}
&T_{\mathrm{c}}(t) =  t_{\mathrm{c}} + A_{\mathrm{b}} \sin( 2\pi f t),\\
&T_{\mathrm{so}}(t) =  t_{\mathrm{so}} + x_{\mathrm{so}}
A_{\mathrm{b}}\sin( 2\pi f t).
\end{split}
\end{equation}
The AC amplitude is $A_{\mathrm{b}}$ and in general
$A_{\mathrm{b}} \ne A_{\mathrm{d}}$. For most calculations we
assume that $x_{\mathrm{so}}=t_{\mathrm{so}}/t_{\mathrm{c}}$, so
at any time the ratio $T_{\mathrm{so}}(t)/T_{\mathrm{c}}(t)$ is a
time independent constant equal to $x_{\mathrm{so}}$. We also
address the case where $x_{\mathrm{so}}\neq
t_{\mathrm{so}}/t_{\mathrm{c}}$, but for simplicity we assume no
phase difference between the tunnel couplings $T_{\mathrm{c}}(t)$
and $T_{\mathrm{so}}(t)$. For the numerical calculations the DD
parameters are taken to be $t_{\mathrm{c}}=0.2$ meV,
$t_{\mathrm{so}}=0.02$ meV, $g_{1}=2$ and $g_{2}=2.4$. The basic
conclusions of this work are general enough and not specific to
these numbers.

\begin{figure}
\includegraphics[width=10.5cm, angle=270]{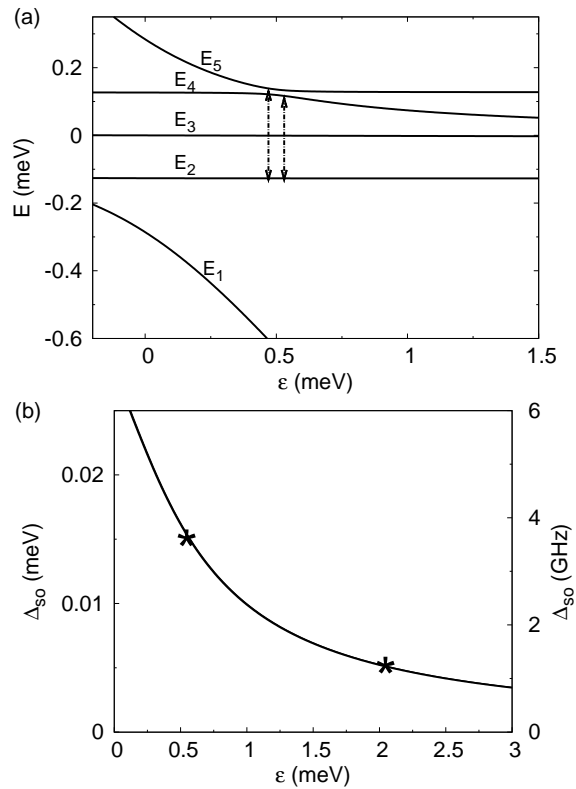}\\
\caption{(a) Two-electron eigenenergies as a function of the
energy detuning for the magnetic field $B=1$ T. The levels
$E_{4}$, $E_{5}$ anticross at $\varepsilon=0.5$ meV due to the
spin-orbit interaction. The two vertical arrows indicate possible
transitions which can be induced by the AC fields defined in the
main text Eq.~(\ref{detuning}) and Eq.~(\ref{barrier}). (b)
Spin-orbit gap ($\Delta_{\mathrm{so}}=E_{5}-E_{4}$ at the
anticrossing point) as a function of the detuning. In this case
the magnetic field is detuning dependent. The AC induced current
is computed for the marked points.}\label{energies}
\end{figure}

\section{Results}

In this section we present the basic results of our work. We
determine the AC induced current for each case of the two AC
fields Eq.~(\ref{detuning}) and Eq.~(\ref{barrier}). The DD
eigenenergies $E_i$ satisfy
$H_{\mathrm{DD}}|\psi_i\rangle=E_i|\psi_i\rangle$ with
$A_{\mathrm{d}}=A_{\mathrm{b}}=0$, and are shown in Fig.~1(a) at
$B=1$ T. When $t_{\mathrm{so}}=0$ and $g_{1}=g_{2}$ singlet and
triplet states are uncoupled. The energy levels $E_{2}$, $E_{3}$
and $E_{4}$ correspond to the pure triplet states $|T_+\rangle$,
$|T_0\rangle$ and $|T_-\rangle$ respectively. These levels are
detuning independent and are Zeeman-split due to the applied
magnetic field. The energy levels $E_{1}$, $E_{5}$ correspond to
pure singlet states which are $|S_{11}\rangle$, $|S_{02}\rangle$
hybridized due to the tunnel coupling $t_{\mathrm{c}}$ and are
independent of the field as can be seen from $H_{\mathrm{DD}}$.
Importantly, the singlet levels $E_{1}$, $E_{5}$ define a
two-level system and for a fixed $t_{\mathrm{c}}$ the
hybridization is controlled by the energy detuning. The two levels
$E_{1}$, $E_{5}$ anticross at $\varepsilon=0$ where the
hybridization is maximum. This is the only anticrossing point in
the energy spectrum for $t_{\mathrm{so}}=0$. However, according to
$H_{\mathrm{DD}}$ when $t_{\mathrm{so}} \ne 0$ the polarized
triplets $|T_\pm\rangle$ couple to the singlet state
$|S_{02}\rangle$. Therefore, as seen in Fig.~1, at
$\varepsilon\approx0.5$ meV the levels $E_{4}$ and $E_{5}$ form an
anticrossing point due to the SOI. Another SOI-induced
anticrossing point is formed at $\varepsilon<0$ between the energy
levels $E_{1}$ and $E_{2}$, but here we consider $\varepsilon>0$
and as in the experiments\cite{ono2017, chorley, xu, marx} we take
$t_{\mathrm{so}}<t_{\mathrm{c}}$.

Because of the SOI ($t_{\mathrm{so}}\ne 0$) and the difference in
the $g$-factors ($g_1 \ne g_2$) the DD eigenstates
$|\psi_{i}\rangle$ are hybridized singlet-triplet states and can
be written in the general form
\begin{equation}
|\psi_{i}\rangle = \alpha_{i}|S_{11}\rangle +
\beta_{i}|T_{+}\rangle + \gamma_{i} |S_{02}\rangle + \zeta_{i}
|T_{-}\rangle + \eta_{i} |T_{0}\rangle.
\end{equation}
The coefficients denoted by Greek letters determine the character
of the states, and are sensitive to the detuning.

One method to probe the SOI anticrossing point is to focus on the
AC frequency range $ 0 < h f \lesssim E_{5} - E_{4}$ and determine
the position of the AC induced current peak. This method has been
theoretically studied in Ref.~\onlinecite{giavaras19a}. Another
method to probe the anticrossing point is to focus on the AC
frequency range $ E_{4} - E_{2} \lesssim h f \lesssim E_{5} -
E_{2}$, and determine the positions of the two AC induced current
peaks. The present work is concerned with the latter method and
the main subject of the present work is to compare the current
peaks induced separately by the two AC fields; the tunnel barrier
modulation and energy detuning modulation. In
Ref.~\onlinecite{ono2017} both methods have been experimentally
investigated under the assumption that the AC field modulates the
energy detuning of the DD. The case where the AC field modulates
simultaneously the interdot tunnel coupling and the energy
detuning might be experimentally relevant,~\cite{nakajima18} but
this case is not pursued in the present work.

In Fig.~1 the SOI anticrossing is formed at
$\varepsilon\approx0.5$ meV for $B=1$ T. A lower magnetic field
shifts the SOI anticrossing point at larger detuning, and the
degree of hybridization due to the SOI decreases. The reason is
that as $\varepsilon$ increases the $|S_{02}\rangle$ character in
the original singlet state ($t_{\mathrm{so}}=0$) is gradually
replaced by the $|S_{11}\rangle$ character. As a result, the SOI
gap $\Delta_{\mathrm{so}}=E_{5}-E_{4}$, defined at the
anticrossing point, decreases with $\varepsilon$ as shown in
Fig.~\ref{energies}(b). For the parameters considered in this
work, the SOI gap can be analytically determined from the
expression~\cite{loss} $\Delta_{\mathrm{so}} = 2
t_{\mathrm{so}}\sqrt{(1-\cos\theta)/2}$, with $\theta = \arctan(
2\sqrt{2}t_{\mathrm{c}}/ \varepsilon)$.

In our previous work~\cite{giavaras19a} we examined the
transitions between the two singlet-triplet states
$|\psi_4\rangle$ and $|\psi_5\rangle$, whose energy levels form
the SOI anticrossing point (Fig.~\ref{energies}). These
transitions give rise to one current peak which is suppressed near
the anticrossing point, in agreement with an experimental
study.~\cite{ono2017} In the present work, we focus on the
transitions between the two pairs of states $|\psi_5\rangle$ and
$|\psi_2\rangle$ as well as $|\psi_4\rangle$ and $|\psi_2\rangle$.
Here, $|\psi_4\rangle$ and $|\psi_5\rangle$ are strongly
hybridized singlet-triplet states, whereas $|\psi_2\rangle$ has
mostly triplet character provided the detuning is large.

\begin{figure}
\includegraphics[width=9.5cm, angle=270]{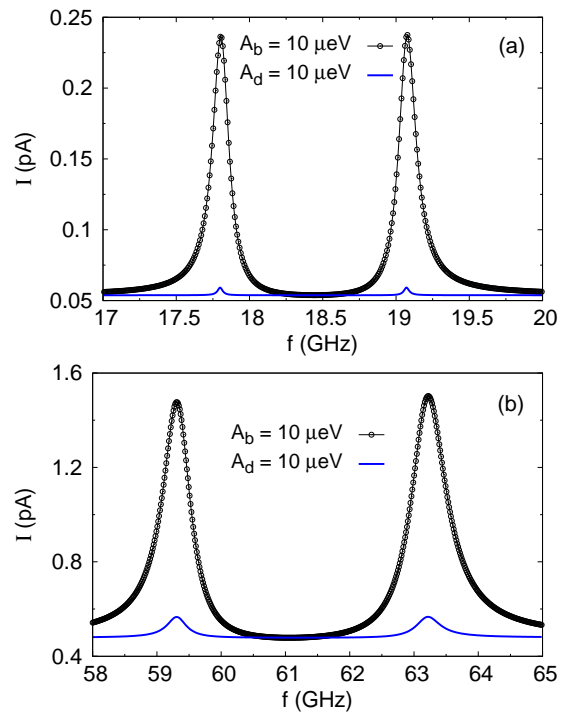}\\
\caption{Current as a function of AC frequency, when the AC field
modulates the tunnel barrier with the AC amplitude
$A_{\mathrm{b}}=10$ $\mu$eV, and the energy detuning with
$A_{\mathrm{d}}=10$ $\mu$eV. The constant value of the detuning is
$\varepsilon=2$, 0.5 meV and the corresponding magnetic field is
$B=0.3$, 1 T for (a) and (b) respectively. These fields define the
singlet-triplet anticrossing point for each value of the
detuning.}\label{ivsf}
\end{figure}

We compute the AC-induced current flowing through the double dot
within the Floquet-Markov density matrix equation of
motion.~\cite{floquet1, floquet2} In this approach we treat the
time dependence of the AC field exactly, taking advantage of the
fact that the DD Hamiltonian is time periodic and thus it can be
expanded in a Fourier series. The model uses for the basis states
of the DD density matrix the periodic Floquet
modes,~\cite{giavaras19b} and consequently it is applicable for
any amplitude of the AC field. In most calculations we take the
parameter $x_{\mathrm{so}}=0.1$ unless otherwise specified.

To study the AC current spectra we choose two values for the
energy detuning $\varepsilon$ (2, 0.5 meV), and determine the
magnetic field at which $|\psi_4\rangle$ and $|\psi_5\rangle$
anticross. At this specific field we plot in Fig.~\ref{ivsf} the
AC-induced current as a function of the AC frequency. As the
energy detuning decreases the magnetic field defining the
corresponding anticrossing point increases. This in turn means
that the AC frequency has to increase to satisfy the corresponding
resonance condition $hf=E_{5}-E_{2}$ (or $hf=E_{4}-E_{2}$). This
increase in the frequency explains the different frequency range
in Fig.~\ref{ivsf}. Furthermore, the off-resonant current is
larger for $\varepsilon=0.5$ meV due to the stronger SOI
hybridization.~\cite{giavaras13}

In the two cases shown in Fig.~\ref{ivsf} two peaks are formed;
one peak is due to the transition between the eigenstates
$|\psi_2\rangle$ and $|\psi_4\rangle$, and the second peak is due
to the transition between $|\psi_2\rangle$ and $|\psi_5\rangle$.
Therefore, the distance between the centres of the two peaks is
equal to the singlet-triplet energy splitting $E_{5}-E_{4}$. For
the specific choice of magnetic field this energy splitting is
equal to the SOI gap of the anticrossing point. For example, for
$\varepsilon=0.5$ meV the gap is $\Delta_{\mathrm{so}} \approx
3.5$ GHz, and for $\varepsilon=2$ meV the gap is
$\Delta_{\mathrm{so}} \approx 1.1$ GHz. These numbers are in
agreement with those derived from the exact energies of the time
independent part of the Hamiltonian $H_{\mathrm{DD}}$.  According
to Fig.~\ref{ivsf}, for a given energy detuning and driving field
the two peaks are almost identical. This is due to the fact, that
at the anticrossing point the states $|\psi_4\rangle$,
$|\psi_5\rangle$ have identical characters when the driving field
is off, and the relevant transition rates are almost equal. In
contrast, away from the anticrossing point the two peaks can be
very different.~\cite{giavaras19b}

\begin{figure}
\includegraphics[width=7.cm, angle=0]{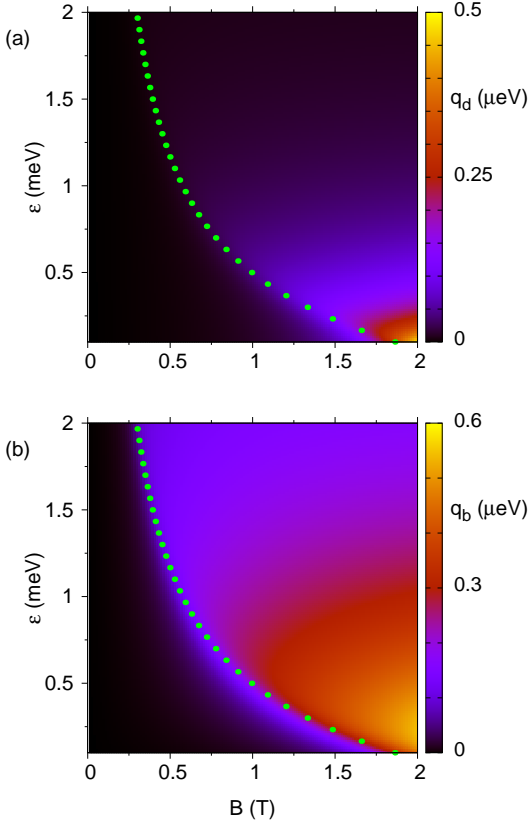}\\
\caption{(a) Absolute value of the coupling parameter
$q_{\mathrm{d}}$ as a function of the energy detuning and magnetic
field for the AC amplitude $A_{\mathrm{d}}=10$ $\mu$eV. The dotted
curve defines the anticrossing point for each $\varepsilon$ and
$B$. (b) The same as (a) but for $q_{\mathrm{b}}$ with
$A_{\mathrm{b}}=10$ $\mu$eV.}\label{coupling}
\end{figure}

\begin{figure}
\includegraphics[width=5.5cm, angle=270]{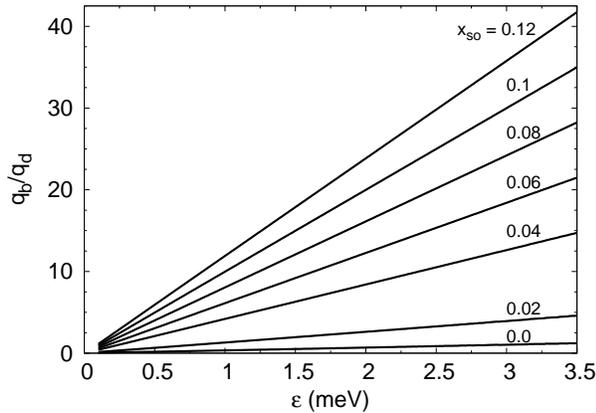}\\
\caption{The ratio $q_{\mathrm{b}}/q_{\mathrm{d}}$ defined in
Eq.~(\ref{ratio}) as a function of detuning for different values
of $x_{\mathrm{so}}$ and
$A_{\mathrm{d}}=A_{\mathrm{b}}$.}\label{xsodet}
\end{figure}

The results in Fig.~\ref{ivsf} demonstrate that the two driving
fields Eq.~(\ref{detuning}) and Eq.~(\ref{barrier}) induce
different peak magnitudes. Specifically, the peaks due to the
tunnel barrier modulation are stronger than those due to the
detuning modulation. As an example, for $\varepsilon=0.5$ meV
[Fig.~\ref{ivsf}(b)] the tunnel barrier modulation induces a
relative peak height of about 1 pA, whereas the relative peak
height is only 0.1 pA for the energy detuning modulation. Some
insight into this interesting behavior can be obtained by
inspecting the time-scale (``Rabi'' frequency) of the coherent
transitions between the eigenstates $|\psi_{2}\rangle$ and
$|\psi_{4}\rangle$. When the AC field modulates the tunnel
coupling, the transitions can be studied within the exact Floquet
eigenvalue problem, but for simplicity we here employ an
approximate approach.~\cite{giavaras19a} This two-level approach
gives the transition frequency $q_{{\mathrm{b}}}/h$, with
\begin{equation}\label{coup}
q_{\mathrm{b}} = \frac{ h f
h^{\mathrm{b}}_{24}}{(h^{\mathrm{b}}_{22} - h^{\mathrm{b}}_{44})}
J_{1}\left(\frac{A_{\mathrm{b}}(h^{\mathrm{b}}_{22} -
h^{\mathrm{b}}_{44})}{hf} \right),
\end{equation}
where $J_{1}(r)$ is a Bessel function of the first kind, and the
argument is $r = A_{\mathrm{b}}(h^{\mathrm{b}}_{22} -
h^{\mathrm{b}}_{44})/hf$, with $h f = E_{4} - E_{2}$ and
\begin{equation}\label{hijb}
\begin{split}
h^{\mathrm{b}}_{ij} =& - \gamma_j ( \sqrt{2}\alpha_i +
x_{\mathrm{so}} \beta_i + x_{\mathrm{so}} \zeta_i )\\ &- \gamma_i
( \sqrt{2}\alpha_j + x_{\mathrm{so}} \beta_j + x_{\mathrm{so}}
\zeta_j ), \quad i, j=2, 4
\end{split}
\end{equation}
When the AC field modulates the energy detuning the time-scale of
the coherent transitions between the eigenstates
$|\psi_{2}\rangle$ and $|\psi_{4}\rangle$ is approximately
$q_{\mathrm{d}}/h$. The coupling parameter $q_{\mathrm{d}}$ is
found by $q_{\mathrm{b}}$ with the replacements
$A_{\mathrm{b}}\rightarrow A_{\mathrm{d}}$ and
$h^{\mathrm{b}}_{ij} \rightarrow h^{\mathrm{d}}_{ij}$, where
\begin{equation}\label{hijd}
\begin{split}
h^{\mathrm{d}}_{ij} =-\gamma_i\gamma_j, \quad i, j=2, 4
\end{split}
\end{equation}
In general, $q_{\mathrm{b}}$ can be very different from
$q_{\mathrm{d}}$, even when $A_{\mathrm{b}}=A_{\mathrm{d}}$.
Therefore, the two driving fields are expected to induce current
peaks with different width and height.

To quantify the two parameters $q_{\mathrm{b}}$, $q_{\mathrm{d}}$
we plot in Fig.~\ref{coupling} $q_{\mathrm{b}}$, $q_{\mathrm{d}}$,
as a function of the energy detuning and the magnetic field. Here,
$A_{\mathrm{d}}=A_{\mathrm{b}}=10$ $\mu$eV, and $h f =
E_{4}-E_{2}$ [in Eq.~(\ref{coup})] is magnetic field as well as
detuning dependent, and is determined by the energies of
$H_{\mathrm{DD}}$. If we denote by $B_{\mathrm{an}}$ the field at
which the anticrossing point is formed, then as seen in
Fig.~\ref{coupling} both $q_{\mathrm{b}}$ and $q_{\mathrm{d}}$ are
large for $B>B_{\mathrm{an}}$, but vanishingly small for $B \ll
B_{\mathrm{an}}$. The reason is that the state $|\psi_{4}\rangle$
is singlet-like for $B>B_{\mathrm{an}}$, but triplet-like for
$B<B_{\mathrm{an}}$, whereas $|\psi_2\rangle$ has mostly triplet
character independent of $B$, provided $\varepsilon$ is away from
zero. Transitions between triplet-like states are in general slow
leading to vanishingly small $q_{\mathrm{b}}$, $q_{\mathrm{d}}$
for $B \ll B_{\mathrm{an}}$. In contrast, if we choose $h f =
E_{5}-E_{2}$, then both $q_{\mathrm{b}}$ and $q_{\mathrm{d}}$ are
large for $B<B_{\mathrm{an}}$. For large enough detuning where the
two spins are in the Heisenberg regime, the exchange energy is
approximately $2 t^{2}_{\mathrm{c}} /\varepsilon$ therefore
$B_{\mathrm{an}}$ satisfies $(g_1+g_2)\mu_{\mathrm{B}}
B_{\mathrm{an}}/2 \approx 2 t^{2}_{\mathrm{c}} /\varepsilon$.

Most importantly Fig.~\ref{coupling} demonstrates that
$q_{\mathrm{b}}>q_{\mathrm{d}}$ when $\varepsilon \gtrsim 0.2$
meV. To understand this result we focus on the anticrossing point
where $r < 1$, then from Eq.~(\ref{coup}) $q_{\mathrm{b}}\approx
h^{\mathrm{b}}_{24} A_{\mathrm{b}}/2$ because $J_{1}(r)\approx
r/2$, and similarly $q_{\mathrm{d}}\approx  h^{\mathrm{d}}_{24}
A_{\mathrm{d}}/2$. Moreover, away from zero detuning the state
$|\psi_2\rangle$ has mostly triplet character, therefore
\begin{equation}
h^{\mathrm{b}}_{24} \approx -\gamma_4 ( \sqrt{2}\alpha_2 +
x_{\mathrm{so}} \beta_2 ) -\gamma_2 ( \sqrt{2}\alpha_4 +
x_{\mathrm{so}} \zeta_4 ),
\end{equation}
and the ratio $q_{\mathrm{b}}/q_{\mathrm{d}}$ is
\begin{equation}\label{ratio}
\frac{ q_{\mathrm{b}} }{ q_{\mathrm{d}} }  =
\frac{A_{\mathrm{b}}}{A_{\mathrm{d}}}\left(\sqrt{2}
\frac{\alpha_2}{\gamma_2} + x_{\mathrm{so}}
\frac{\beta_2}{\gamma_2} +  \sqrt{2} \frac{\alpha_4}{\gamma_4} +
x_{\mathrm{so}} \frac{\zeta_4}{\gamma_4}\right).
\end{equation}
As $\varepsilon$ increases $\beta_2 \rightarrow 1$, $\gamma_2\ll
1$ and, considering absolute values, the second term in
Eq.~(\ref{ratio}) dominates
\begin{equation}
\frac{\beta_2}{\gamma_2} \gg \frac{\alpha_2}{\gamma_2},
\frac{\alpha_4}{\gamma_4}, \frac{\zeta_4}{\gamma_4}.
\end{equation}
Consequently, $q_{\mathrm{b}}$ can be much greater than
$q_{\mathrm{d}}$, especially at large $\varepsilon$, and for a
fixed tunnel coupling $t_{\mathrm{c}}$ the exact value of the
ratio $q_{\mathrm{b}}/q_{\mathrm{d}}$ depends sensitively on
$x_{\mathrm{so}}$. This demonstrates the importance of the time
dependence of the spin-orbit coupling. The conclusions derived
from the parameters $q_{\mathrm{b}}$, $q_{\mathrm{d}}$ assume that
there is no `multi-level' interference and only the levels
$E_{i}$, $E_{j}$ satisfying $hf=|E_{i}-E_{j}|$ are responsible for
the current peaks. The approximate results are more accurate when
the argument $r$ of the Bessel function is kept small.

To examine the $x_{\mathrm{so}}$-dependence, we consider
$A_{\mathrm{b}}=A_{\mathrm{d}}$ and plot in Fig.~\ref{xsodet} the
ratio $q_{\mathrm{b}}/q_{\mathrm{d}}$ versus the detuning at the
anticrossing point, and for different values of $x_{\mathrm{so}}$.
By increasing $\varepsilon$ and for large values of
$x_{\mathrm{so}}$ the coupling parameters $q_{\mathrm{b}}$,
$q_{\mathrm{d}}$ can differ by over an order of magnitude;
$q_{\mathrm{b}}/q_{\mathrm{d}}>10$. This leads to (very) different
current peaks with the tunnel barrier modulation inducing stronger
peaks. The special value $x_{\mathrm{so}}=0$ corresponds to a time
independent SOI tunnel coupling [see Eq.~(\ref{barrier})], and the
special value $x_{\mathrm{so}}=0.1$ corresponds to a time
independent ratio $T_{\mathrm{so}}/T_{\mathrm{c}}=0.1$. Although,
the ratio $q_{\mathrm{b}}/q_{\mathrm{d}}$ can be computed at any
$\varepsilon$, the regime of small $\varepsilon$ ($<0.2$ meV) is
not particularly interesting in this work. The reason is that with
decreasing $\varepsilon$ the character of the state
$|\psi_2\rangle$ changes from triplet-like to singlet-triplet,
which eventually becomes approximately equally populated to
$|\psi_4\rangle$ and $|\psi_5\rangle$. Therefore, the current
peaks induced by both driving fields are suppressed even when
$q_{\mathrm{b}}$ or $q_{\mathrm{d}}$ is large. In
Fig.~\ref{xsodet} the maximum value of the detuning is chosen to
give $\varepsilon/t_{\mathrm{c}}\approx 17.5$ which can be easily
achieved in double quantum dots. Some experiments~\cite{petta,
ono2017, perge} have reported values greater than
$\varepsilon/t_{\mathrm{c}}\approx 100$, thus $q_{\mathrm{b}}$ can
be even two orders of magnitude greater than $q_{\mathrm{d}}$.

\begin{figure}
\includegraphics[width=9.5cm, angle=270]{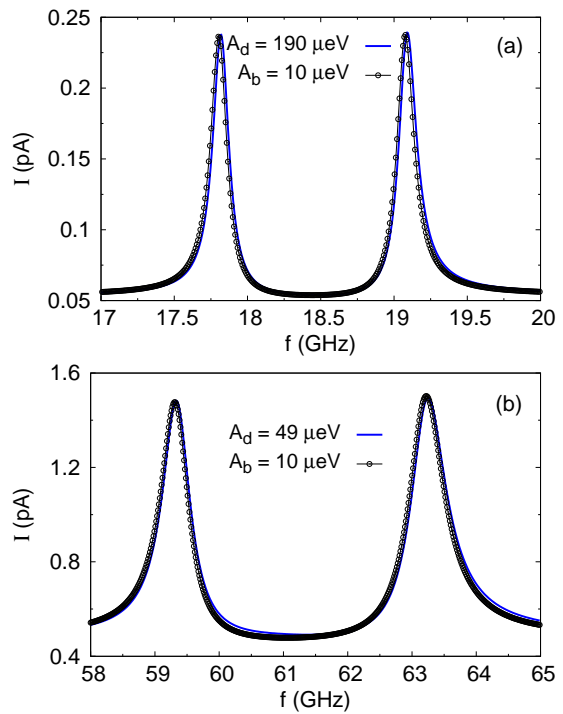}\\
\caption{As in Fig.~\ref{ivsf}, but (a) $A_{\mathrm{b}}=10$
$\mu$eV and $A_{\mathrm{d}}=19A_{\mathrm{b}}$, (b)
$A_{\mathrm{b}}=10$ $\mu$eV and
$A_{\mathrm{d}}=4.9A_{\mathrm{b}}$. The value of $A_{\mathrm{d}}$
is chosen so that to approximately induce the same current peaks
as those induced by $A_{\mathrm{b}}$.}\label{compa}
\end{figure}

According to the above analysis if
$q_{\mathrm{b}}/q_{\mathrm{d}}\approx 1$ then the current peaks
induced by the two driving fields should approximately display the
same characteristics. As an example, consider the two sets of
current peaks shown in Fig.~\ref{ivsf} both for
$x_{\mathrm{so}}=0.1$ and $\varepsilon=2$ meV, $\varepsilon=0.5$
meV respectively. Focusing on $x_{\mathrm{so}}=0.1$ in
Fig.~\ref{xsodet}, we see that at $\varepsilon=2$ meV
$q_{\mathrm{b}}/q_{\mathrm{d}}\approx 19$ and at $\varepsilon=0.5$
meV $q_{\mathrm{b}}/q_{\mathrm{d}}\approx 4.9$. These numbers
suggest that if at $\varepsilon=2$ meV we choose for the AC
amplitudes the ratio $A_{\mathrm{b}}/A_{\mathrm{d}}\approx 1/19$
then the detuning and the barrier modulation should induce
approximately the same peak characteristics. Likewise at
$\varepsilon=0.5$ meV the ratio should be
$A_{\mathrm{b}}/A_{\mathrm{d}}\approx 1/4.9$. These arguments are
quantified in Fig.~\ref{compa} where we plot the current peaks for
the two driving fields for different AC amplitudes satisfying the
condition $q_{\mathrm{b}}/q_{\mathrm{d}}\approx 1$. The results
confirm that the induced current peaks display approximately the
same characteristics.

\begin{figure}
\includegraphics[width=5.5cm, angle=270]{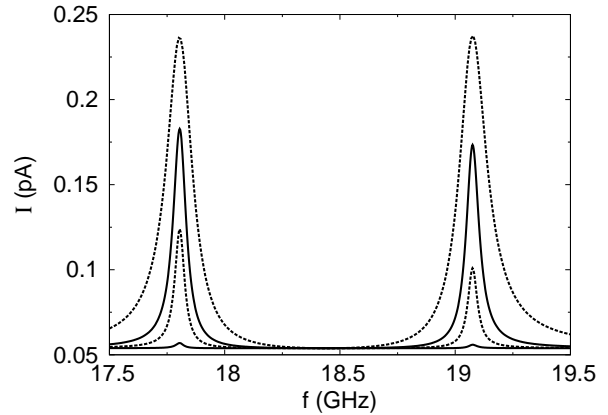}\\
\caption{Current as a function of AC frequency, when the AC field
modulates the tunnel barrier, with the AC amplitude
$A_{\mathrm{b}}= 10$ $\mu$eV. The detuning is $\varepsilon=2$ meV
and from the upper to the lower curve the parameter
$x_{\mathrm{so}}=0.1$, 0.04, 0.02, 0.}\label{xsopeaks}
\end{figure}

Inducing strong current peaks can be advantageous in order to
perform spectroscopy of the singlet-triplet levels and extract the
SOI anticrossing gap. However, an important aspect is that the SOI
gap cannot be extracted from the positions of the current peaks at
arbitrary large AC amplitudes. In particular, by increasing the AC
amplitude the two peaks start to overlap and eventually the
resonant pattern of the current changes
drastically.~\cite{giavaras19b} Therefore, the distance between
the two peaks cannot accurately predict the SOI gap. This effect
has been theoretically studied for the case of a time dependent
energy detuning,~\cite{giavaras19b} and it can be readily shown
that similar trends occur for a time dependent tunnel coupling.
The driving regime where the two current peaks strongly overlap is
not considered in the present work, since it is not appropriate
for the spectroscopy of the SOI gap.

Finally, in Fig.~\ref{xsopeaks} we plot the current peaks when the
AC field modulates the tunnel barrier with the amplitude
$A_{\mathrm{b}}=10$ $\mu$eV and the constant detuning
$\varepsilon=2$ meV. With decreasing $x_{\mathrm{so}}$ the two
peaks gradually weaken and for $x_{\mathrm{so}}=0$ the peaks are
vanishingly small; for this value the peaks are of the same order
as the peaks induced by the detuning modulation with the same
amplitude $A_{\mathrm{d}}=10$ $\mu$eV (for clarity these peaks are
not shown). The small difference between the left and the right
peaks, for example when $x_{\mathrm{so}}=0.02$, can be understood
by inspecting the different values of $q_{\mathrm{b}}$
[Eq.~(\ref{coup})] which involve different matrix elements and
frequencies. The overall trends indicate the important role of the
time dependent spin-orbit term and are consistent with the results
shown in Fig.~\ref{xsodet}. As $x_{\mathrm{so}}$ decreases the
coupling parameter $q_{\mathrm{b}}$ decreases too, thus the time
scale of the singlet-triplet transitions becomes longer leading to
smaller peaks. Moreover, by decreasing $x_{\mathrm{so}}$,
$q_{\mathrm{b}}$ becomes approximately equal to $q_{\mathrm{d}}$,
therefore the tunnel barrier modulation and the detuning
modulation result in approximately the same current peaks.

\section{Summary}

In summary, we considered a double quantum dot in the spin
blockade regime and studied the AC induced current peaks for a
specific energy configuration which involves two hybridized
singlet-triplet states as well as a third state with mostly
triplet character. The two AC induced transitions which rely on
the spin-orbit interaction, result in two current peaks. We found
that for a large energy detuning the two peaks are stronger when
the time periodic field modulates the interdot tunnel coupling
(barrier) instead of the energy detuning. We demonstrated that a
time dependence in the spin-orbit coupling can significantly
modify the peak characteristics, and should be taken into account
even when the actual spin-orbit coupling is small. Our work
suggests an efficient way of probing the spin-orbit energy gap in
two-spin states based on transport measurements.

\setcounter{secnumdepth}{0} 

\section{Acknowledgement}

Part of this work was supported by CREST JST (JPMJCR15N2), and by
JSPS KAKENHI (18K03479).




\end{document}